\documentclass[twocolumn,aps,prl,amsmath,amssymb,amsfonts,superscriptaddress,notitlepage]{revtex4-1}
\usepackage{graphicx}  
\usepackage{dcolumn}   
\usepackage{bm}        
\usepackage{amssymb}   
\usepackage[svgnames]{xcolor}
\usepackage[colorlinks=true,
            linkcolor=red,
            urlcolor=blue,
            citecolor=DarkGreen]{hyperref}

\begin{document}

\title{Single-Shot Quantum Non-Demolition Detection of Itinerant Microwave Photons}
\author{Jean-Claude~Besse}\email{jbesse@phys.ethz.ch}
\author{Simone~Gasparinetti}
\author{Michele~C.~Collodo}
\author{Theo~Walter}
\author{Philipp~Kurpiers}
\author{Marek~Pechal}\thanks{Current address: Ginzton Laboratory, Stanford University, Stanford, California 94305, USA}
\author{Christopher~Eichler}
\author{Andreas~Wallraff}\affiliation{Department of Physics, ETH Zurich, CH-8093 Zurich, Switzerland}

\date{\today}

\begin{abstract}
Single-photon detection is an essential component in many experiments in quantum optics, but remains challenging in the microwave domain.
We realize a quantum non-demolition detector for propagating microwave photons and characterize its performance using a single-photon source. To this aim we implement a cavity-assisted conditional phase gate between the incoming photon and a superconducting artificial atom. By reading out the state of this atom in single shot, we reach an internal photon detection fidelity of 71\%, limited by the coherence properties of the qubit. By characterizing the coherence and average number of photons in the field reflected off the detector, we demonstrate its quantum non-demolition nature.
We envisage applications in generating heralded remote entanglement between qubits and for realizing logic gates between propagating microwave photons.
\end{abstract}

\maketitle

Single-photon detectors~\cite{Hadfield2009} for itinerant fields are a key element in remote entanglement protocols~\cite{Chou2005}, in linear optics quantum computation~\cite{Kok2007,Knill2001}, and in general to characterize correlation properties of radiation fields~\cite{Glauber1963b}. While such detectors are well established at optical frequencies, their microwave equivalents are still under development, partly due to the much lower photon energy in this frequency band~\cite{Gu2017}.
At microwave frequencies, itinerant fields are typically recorded with linear detection schemes, analogous to optical homodyne detection~\cite{Eichler2012}. Such detection can now be realized with high efficiency by employing near quantum limited parametric amplifiers~\cite{Castellanos2007} and furthermore allows for a full tomographic characterization of radiation fields \cite{Bozyigit2011}. However, protocols such as entanglement heralding require the intrinsic nonlinearity of a single-photon detector in order to yield high purity states despite losses between the source and the detector. Such a component has therefore raised interest in the community, leading to a variety of theoretical proposals~\cite{Romero2009,Helmer2009b,Peropadre2011,Witthaut2012,Manzoni2014,Fan2014a,Sathyamoorthy2014,Koshino2015, Sathyamoorthy2016b,Kyriienko2016,Koshino2016a} as well as initial experimental demonstrations in the last decade~\cite{Gleyzes2007,Johnson2010,Leek2010,Chen2011a,Inomata2016,Narla2016,Kono2017a}.

The first photodetection experiment in the superconducting regime that did not require photons to be stored in high-quality factor cavities~\cite{Gleyzes2007,Johnson2010,Leek2010} was based on current biased Josphson junctions~\cite{Chen2011a}, but was destructive and involved a long dead time. Later, systems involving absorption into artificial atoms (and thus destruction) of traveling photons~\cite{Inomata2016,Narla2016} were implemented. Very recently single photons have been heralded from a non-demolition detection scheme based on a photon-qubit entangling gate, similar in spirit to this manuscript, using a strong dispersive shift in a 3D cavity~\cite{Kono2017a}.

Here we demonstrate single-shot quantum non-demolition (QND) detection of itinerant single photons in the microwave domain, based on a cavity-assisted controlled phase gate between an artificial atom and a propagating photon~\cite{Duan2004}. We show the unconditional detection of an itinerant photonic wave packet containing a Fock state at the single photon level.

\begin{figure}
\includegraphics{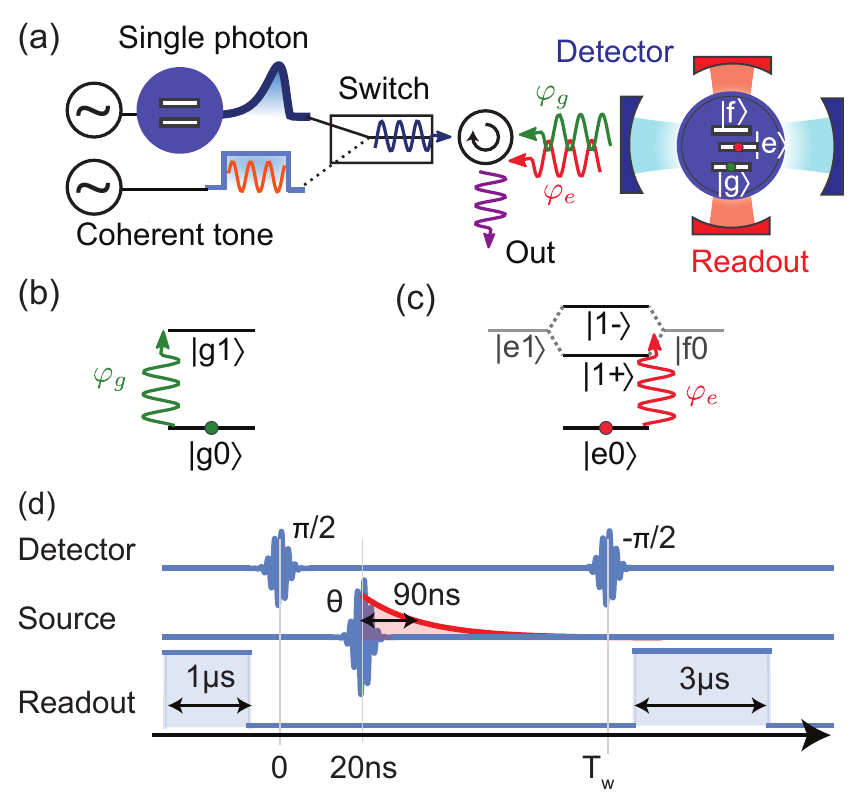}
\caption{\label{fig:jaynes_phase} Principle of quantum non-demolition single-photon detection. (a) Sketch of the setup connecting a source of single photons or a coherent source via an on-chip switch and a circulator to the input of the detector. (b,c) Energy-level diagram of the atom-cavity system when the atom is either in the ground (b) or in the excited state (c), contrasting the harmonic ladder of a bare cavity in (b) to the Jaynes-Cummings anharmonic ladder in (c).
(d) The pulse scheme consisting of a $\pi/2$ and $-\pi/2$ pulses on the detection-qubit, defining the length of the time window $T_\mathrm{w}$, as well as a pulse on the source-qubit of Rabi amplitude $\theta$. The emitted photon lineshape is sketched in red. A readout pulse is used to measure the state of the detection-qubit at the end of the protocol, as well as to preselect the single shot traces to discard thermal population.
}
\end{figure}

Our setup consists of a transmon-type superconducting artificial atom coupled to two resonators acting as single-mode cavities, see Fig.~\ref{fig:jaynes_phase}a for a sketch and Suppl.~Mat.~for details. Both resonators are coupled to a Purcell filter~\cite{Reed2010} to protect the qubit from decay into the output lines. We tune the first to second-excited state transition of the transmon, $\omega_\mathrm{ef}/(2\pi)=6135$~MHz, into resonance with the \emph{detector} resonator. The transmon anharmonicity is $\alpha/(2\pi)=-340$~MHz, such that its ground to first-excited state transition is at $\omega_\mathrm{ge}/(2\pi)=6475$~MHz. The transmon is coupled to the detector resonator with rate $g_\mathrm{0}/(2\pi)=40$~MHz and the latter has an effective linewidth $\kappa/(2\pi)=19$~MHz. The resulting level diagram is displayed in Fig.~\ref{fig:jaynes_phase}b,c. With the transmon in the ground state (panel b), photons impinging on the detector at the resonator frequency acquire a phase $\varphi_g=\pi$ as they are reflected. By contrast, with the transmon in the first excited state (panel c), photons of the same frequency are reflected without interacting with the cavity, and thus acquire no phase ($\varphi_e=0$).
The \emph{readout} resonator, at $\omega_\mathrm{ro}/(2\pi)=4800$~MHz, is used to perform high-fidelity, dispersive single-shot readout of the qubit state~\cite{Walter2017}.
We connect the input of our detector with the output to a single photon source embedded in an on-chip switch, see Ref.~\cite{Pechal2016}.
The single photon source is a transmon strongly coupled to its output port with an emission linewidth of $1.77~\rm{MHz}$.
The switch is based on a combination of hybrid couplers and tunable resonators. It enables toggling between its two inputs, supplying either a coherent tone from a conventional microwave generator or a single-photon wave packet emitted by the source-qubit to the input of the detector.

Our protocol for single photon detection (Fig.~\ref{fig:jaynes_phase}d) begins with a measurement of the transmon state, in order to reject those instances in which the qubit is found to be thermally excited (6\% of the total traces were discarded, see Suppl.~Mat.~for details). We then prepare the transmon in the superposition state $(|g\rangle+|e\rangle)/\sqrt{2}$ with a $\pi/2$ pulse. This defines the time $t=0$, at which the detection window of length $T_\mathrm{w}$ begins. 20~ns later we emit a photon wave packet in the state $|\gamma\rangle = \cos(\theta/2) \left| 0 \right\rangle + \sin(\theta/2) \left| 1 \right\rangle$, a coherent superposition of vacuum and a single-photon Fock state, with a relative weight set by the preparation angle $\theta$ of the Rabi pulse applied to the source-qubit.
At time $t=T_\mathrm{w}$, we apply a $-\pi/2$ pulse to the transmon, effectively completing a Ramsey sequence, and immediately measure the qubit state.

\begin{figure}
\includegraphics{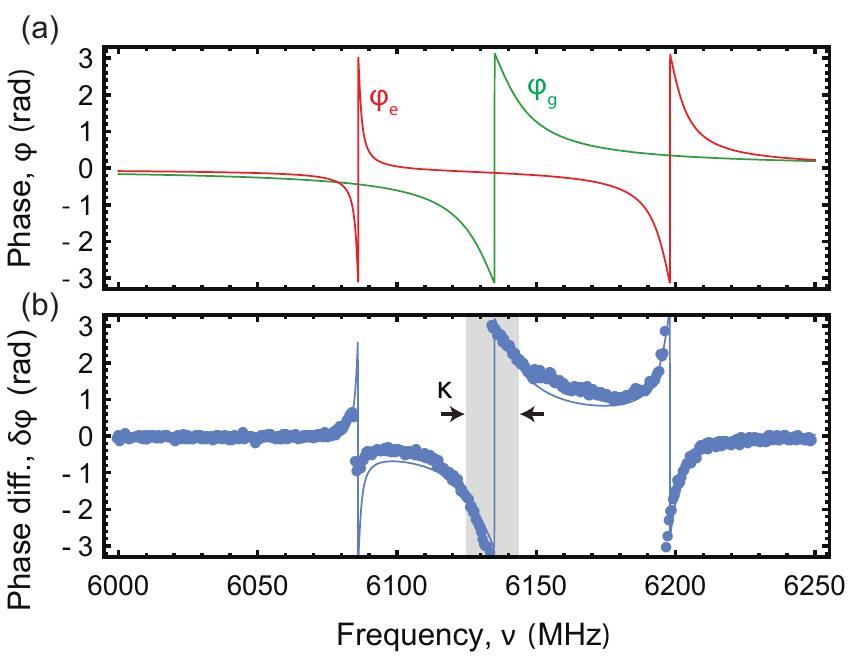}
\caption{\label{fig:detection_pulsescheme} (a) Expected phase $\varphi_\mathrm{g}$ ($\varphi_\mathrm{e}$) of a weak coherent signal upon reflection off of the cavity-atom system, when the detection-qubit is in $|g\rangle$ (green) ($|e\rangle$ (red)). (b) Phase difference $\delta\varphi$ measured in pulsed spectroscopy (blue dots), with model (solid line). The linewidth $\kappa$ of the bare cavity is indicated by the shaded area.}
\end{figure}

We first characterize the response of the detector by considering the phases $\varphi_\mathrm{e,g}$ acquired by a weak coherent tone reflected off the detector input, dependent on the state of the transmon (Fig.~\ref{fig:detection_pulsescheme}a,b). We measure the difference $\delta\varphi=\varphi_\mathrm{g}-\varphi_\mathrm{e}$ by pulsed spectrosopy and find $\delta\varphi=\pi$ at the cavity frequency $\omega_\mathrm{ef}$, as well as at the dressed-state frequencies $\omega_\mathrm{ef}\pm \sqrt{2}g_\mathrm{0}$. In these configurations, a controlled-phase gate is realized between the qubit and a propagating photon. For a definite phase to be acquired by the photon, its spectral bandwidth is required to be smaller than the detector cavity linewidth, $\kappa$. With that condition fulfilled, the gate is not dependent on the temporal shape of the photon.

\begin{figure}[!ht]
\includegraphics{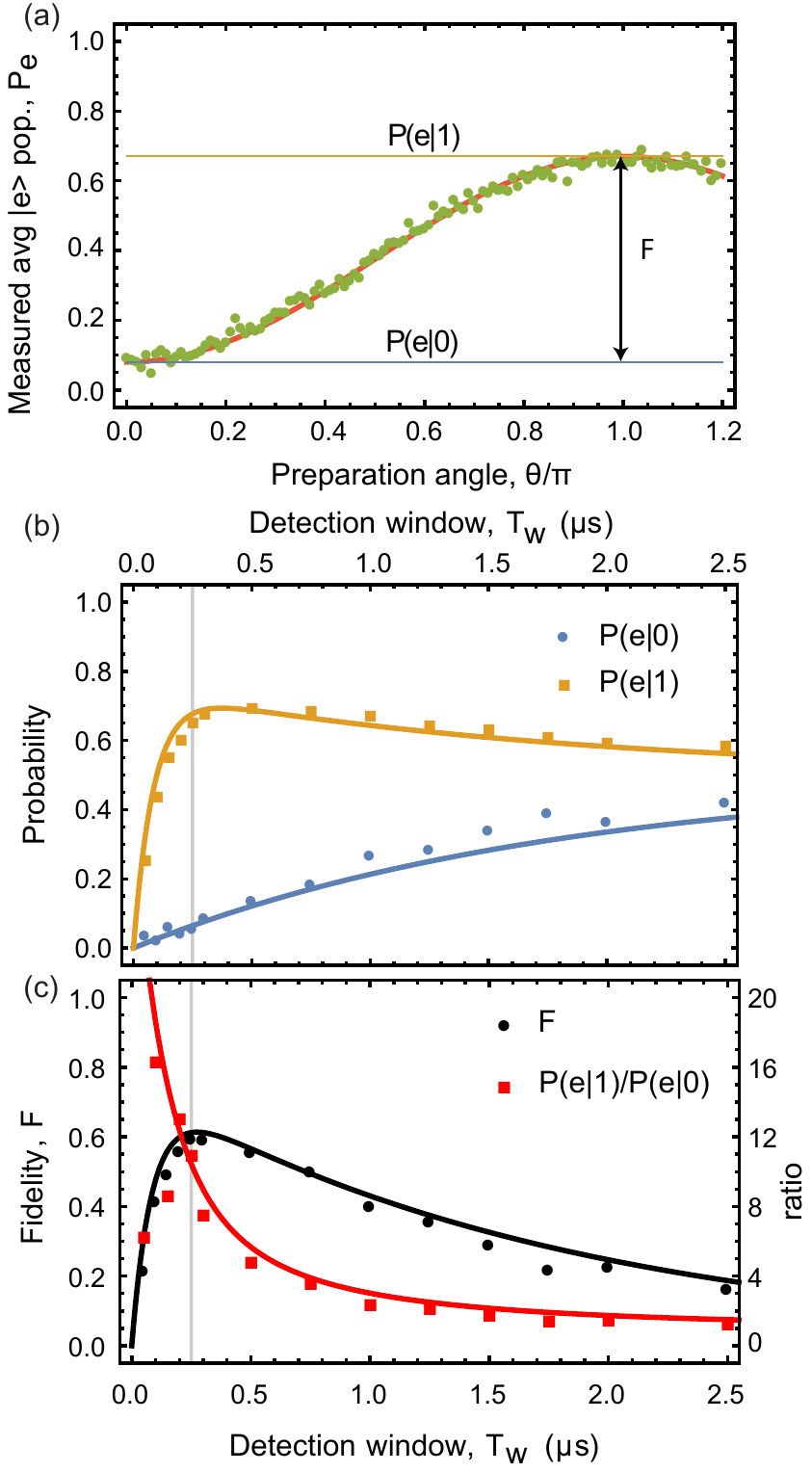}
\caption{\label{fig:timewindow}  (a) Measured (averaged readout) excited state population of the detection-qubit (green dots) as a function of the preparation angle $\theta$ of the photon state $|\gamma\rangle$, for $T_\mathrm{w}=250$~ns. The red line is a fit to the expected dependence, with the detection efficiency $P(e|1)=65.8\%$ (orange) and the dark count probability $P(e|0)=5.9\%$ (blue line) as parameters. The fidelity $F=59.9\%$ is indicated. (b) Dark count probability $P(e|0)$ (blue), photon detection efficiency $P(e|1)$ (yellow), (c) their ratio $P(e|1)/P(e|0)$ (red), and the fidelity $F$ (black) vs. length of the detection time window $T_\mathrm{w}$. Theory lines take into account finite lifetime, losses and photon lineshape (see main text). The length of detection time window $T_\mathrm{w}=250$~ns is used in the remainder of the paper (vertical gray line).}
\end{figure}

The Ramsey sequence displayed in Fig.~\ref{fig:jaynes_phase}d realizes a phase gate on the detection-qubit controlled by the presence of the photon. This process implements the photon detection. To test the fidelity of this protocol with single photons at the input, we measure the average excited state population of the detection-qubit as a function of the preparation angle of the photon source $\theta$ (Fig.~\ref{fig:timewindow}a, $T_\mathrm{w}=250$~ns). The scaling factor between pulse amplitude and preparation angle for the source is independently calibrated in a Rabi experiment.
The data follows a sine-squared dependence, corresponding to the average photon number prepared by the source, with a reduced visibility characterized by the probability $P(e|1)=65.8\%$ of measuring the detection-qubit in the excited state when a photon is emitted and the probability $P(e|0)=5.9\%$ of measuring the detection-qubit in the excited state without emitting a photon. In the context of photon detection we refer to $P(e|1)$ as the detection efficiency and $P(g|0)$ as the dark count probability. As a performance metric we define the detection fidelity $F=1-P(g|1)-P(e|0)=P(e|1)-P(e|0)=59.9\%$ as the difference between detection efficiency and dark count probability.

To gain insight into the sources of errors in the protocol, we extract the detection efficiency and dark count probability vs.~the length of detection time window, $T_\mathrm{w}$ (Fig.~\ref{fig:timewindow}b).
The detection efficiency peaks at an optimal length of $T_\mathrm{w}\simeq 300$~ns, while the dark count probability monotonically increases. A first source of errors is the limited coherence $T_2^*=1.8\,\mu$s of the detection-qubit in a Ramsey experiment. A second type of error is due to loss in the components which connect the source to the detector. The measured total loss in the switch, the coaxial cables and the circulator was found to be approximately 25\% (see Suppl.~Mat.~for calibration measurements using the nonlinear response of the source and the detector). As a result, approximately 75\% of the photons emitted by the source reach the detector, leading to an overall scaling factor independent of $T_\mathrm{w}$. Finally, for short detection windows, part of the photon envelope is cut off by terminating the protocol with the $-\pi/2$ pulse, limiting the detection efficiency. The trends in Fig.~\ref{fig:timewindow}b (solid lines) are quantitatively explained by three sources of errors which we have characterized independently.
For entanglement distribution and other heralded experiments, the ratio $P(e|1)/P(e|0)$ between detection efficiency and dark count probability directly relates to the error rate. We report this ratio, together with the fidelity $F$ vs. the length of detection time window $T_\mathrm{w}$, in Fig.~\ref{fig:timewindow}c. While the fidelity peaks at around the same $T_\mathrm{w}$ as the detection efficiency, the ratio still improves for shorter $T_\mathrm{w}$ as the dark count probability approaches zero. In our case, this ratio reaches up to 16, with significant variations at short $T_\mathrm{w}$ attributed to fluctuations in the low dark count probability.

The fidelity extracted from Fig.~\ref{fig:timewindow}a refers to an averaged readout. When performing single-shot readout in $100$~ns (see Suppl.~Mat.), we find that the total fidelity of detecting single photons for $T_\mathrm{w}=250$~ns is $F=50\%$. This value agrees with the one obtained from the averaged measurements after taking into account the measured 92\% readout fidelity, mainly limited by the transmon decay time $T_1=3.0~\mu$s.
The infidelity is due to imperfect detection efficiency, $P(g|1)=37\%$, and dark count probability, $P(e|0)=13\%$. After accounting for the calibrated losses before the detector, the internal probability to miss a photon is $P_\mathrm{in}(g|1)=16\%$ (corresponding to a quantum detection efficiency of 0.84), so our detector has an internal fidelity of $F_\mathrm{in}=1-P_\mathrm{in}(g|1)-P(e|0)=71\%$.

\begin{figure}
\includegraphics{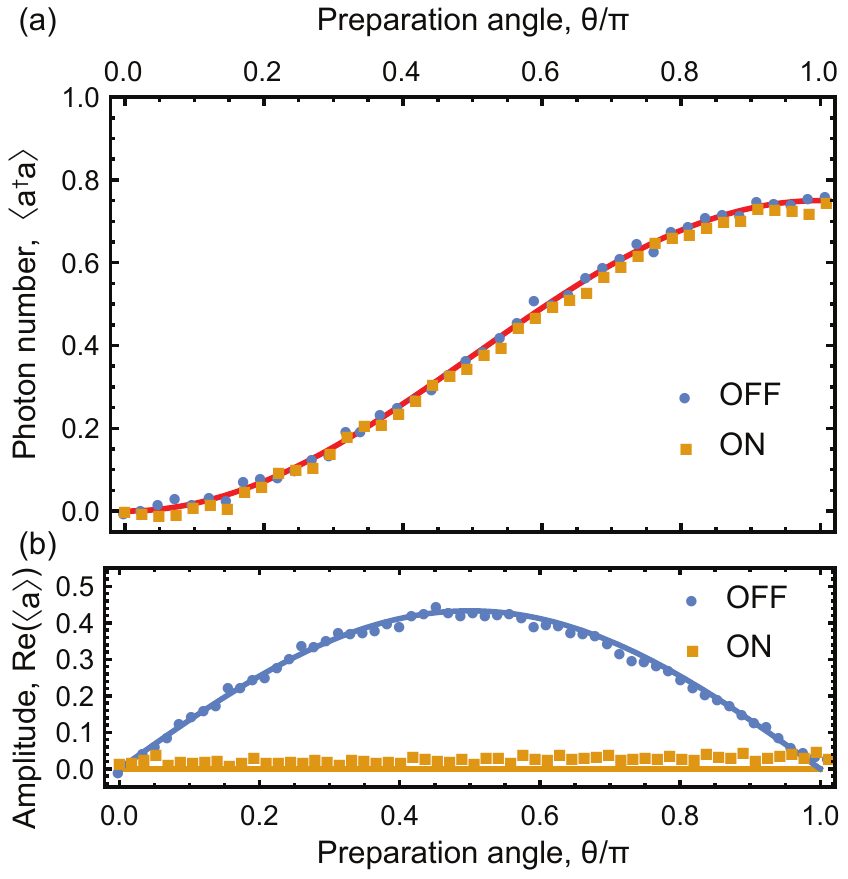}
\caption{\label{fig:QND} Non-demolition character of the measurement. (a) Power, reported as photon number $\langle a^\dagger a\rangle$ is conserved while (b) phase, measured as optimized quadrature $\mathrm{Re}(\langle a \rangle)$, is erased. Solid lines are the expected response of a QND detector.}
\end{figure}

To test the quantum non-demolition nature of the photon detection, we employ a linear amplification chain to measure the average photon number and coherence of the radiation field reflected off the detector. We consider two states of the detector. The ``ON'' state describes the operation reported up to here. In the ``OFF'' state, we keep the atom in the ground state and detune its frequency to be far off-resonant from the cavity. In both cases, the source emits the same radiation field.
We monitor the ensemble averaged photon number ($\langle a^\dagger a \rangle$ moment) and amplitude ($\text{Re}(\langle a \rangle)$ moment, in the optimized quadrature) by integrating the time traces with a filter matched to the temporal shape of the photon~\cite{Eichler2011}. The results are reported in Fig.~\ref{fig:QND}, together with a model. We scale the axes globally by the separately calibrated loss.
Up to our measurment accuracy of 2\%, we observe no difference in power whether the detection pulse sequence is executed or not (Fig.~\ref{fig:QND}a). Accordingly, we conclude that we are performing a quantum non-demolition measurement. However, the phase of the incoming photon state $|\gamma\rangle$ is randomized (Fig.~\ref{fig:QND}b), as quantum mechanics imposes for the conjugate variables $\{n,\phi\}$. We measure a small remaining coherence offset in the ON measurement, even without emitting any photons ($\theta=0$). We ascribe this offset to unintended driving of the $e$-$f$ transition by the first Ramsey pulse, resulting in the subsequent emission of phase-coherent radiation at the frequency $\omega_\mathrm{ef}$.

The phase difference of $\delta\varphi=\pi$ occurs for any Fock state with $n>0$ at the cavity frequency, such that we expect our detector to click for any wave packet containing $n>0$ photons. Detecting photons at one of the dressed state frequencies in the $n$\textsuperscript{th} manifold, and taking advantage of the photon-blockade effect generated by the Jaynes-Cummings ladder could yield an operation mode that projectively selects the Fock state with $n$ photons. This could be useful in entanglement schemes where a particular Bell state is associated to a definite photon number $n$.

We note that in principle, the protocol can be run continuously with a dead time on the order of the single shot readout time of 100~ns by using the readout result as the initial state for the next iteration. One does not need to perform active feedback, nor discard the results of an initially excited atom, but instead could simply invert the association between the measured qubit state and the presence of a photon.

A clicking detector for itinerant photons in the microwave regime, independent of their temporal shape and with internal fidelity limited by qubit coherence, adds to the circuit QED toolbox for characterization of propagating quantum radiation fields.
We have demonstrated single photon detection with radiation fields at the quantum level, composed of a superposition of vacuum and an $n=1$ photon Fock state. Our device does not internally lose photons upon detection and is built with separate detection and readout lines, which provides easy access to the reflected radiation field. This allows to take advantage of the non-demolition nature of the detector and use the device as a mediator of photon-photon interactions for all photonic quantum computation~\cite{Hacker2016b,Kokkoniemi2017,Koshino2016,Wang2016a,Hua2015}.
Other applications include heralded entanglement~\cite{Narla2016} with high rate without the need to shape the photons, or to perform Bell state analysis~\cite{Witthaut2012}.

\section{Acknowledgments}
The authors thank Alexandre Blais, Christian Kraglund Andersen and Paul Magnard for useful discussions. This work is supported by the European Research Council (ERC) through the “Superconducting Quantum Networks” (SuperQuNet) project, by the National Centre of Competence in Research “Quantum Science and Technology” (NCCR QSIT), a research instrument of the Swiss National Science Foundation (SNSF), and by ETH Zurich.

\begingroup
\def\refname{References}
\def\bibname{References}

\endgroup

\clearpage

\setcounter{equation}{0}
\setcounter{enumiv}{0}
\setcounter{figure}{0}
\setcounter{table}{0}
\makeatletter
\renewcommand{\theequation}{S\arabic{equation}}
\renewcommand{\thefigure}{S\arabic{figure}}
\renewcommand{\bibnumfmt}[1]{[#1]}
\renewcommand{\citenumfont}[1]{#1}
\section{Supplemental Material}
\subsection{Sample fabrication and cabling}
\begin{figure}[b]
\includegraphics[width=\columnwidth]{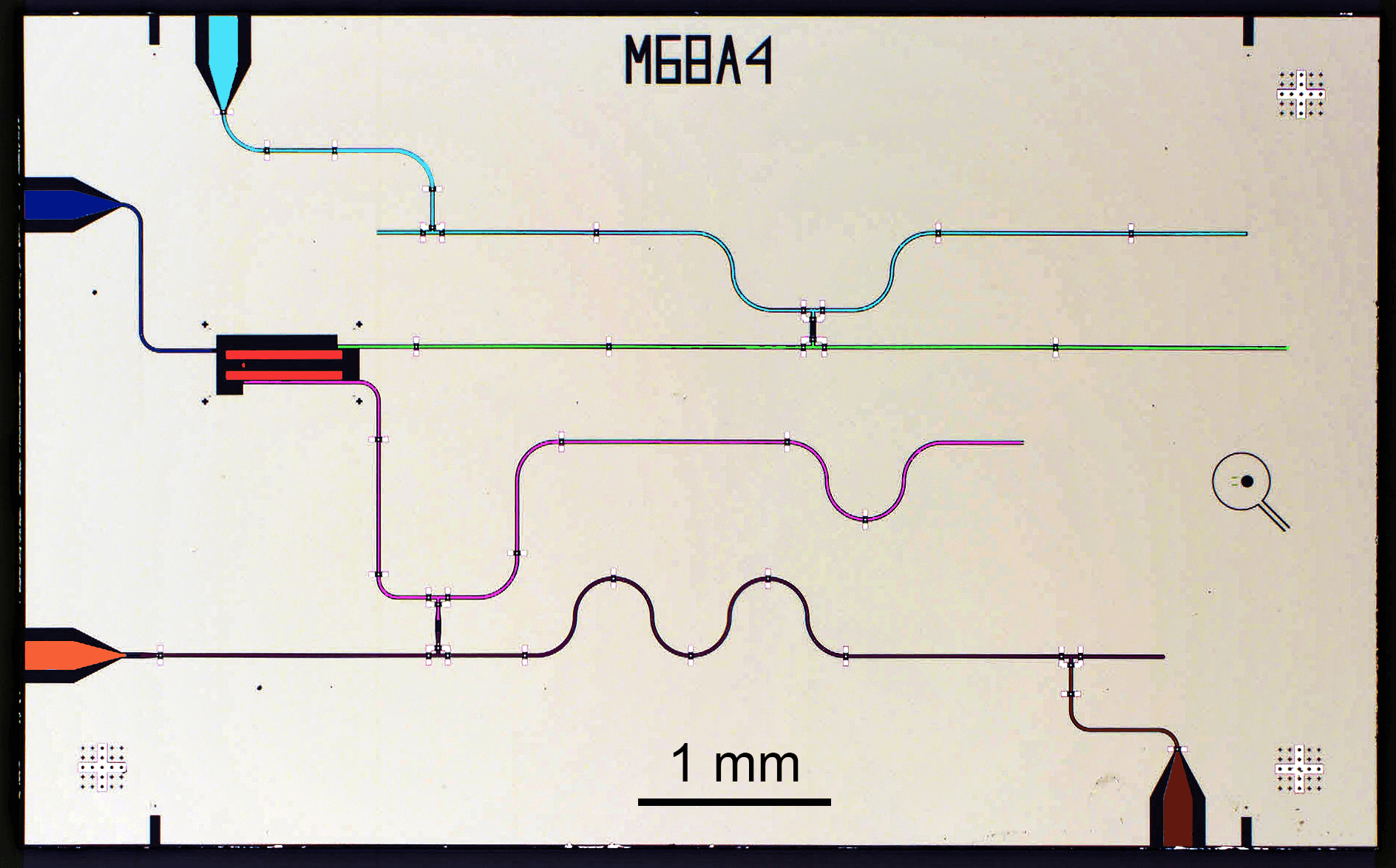}
\caption{\label{fig:samples} False color micrograph of the detector sample. A transmon qubit (red) is coupled to the detection cavity (green) and its Purcell filter (light blue), as well as to a readout cavity (purple) and an according Purcell filter (brown). A charge line (dark blue) allows for driving the qubit and a weakly coupled input port (orange) allows for transmission measurements through the readout resonator. Sapphire is shown in dark and Niobium in light gray.}
\end{figure}

\begin{figure}
\includegraphics[width=\columnwidth]{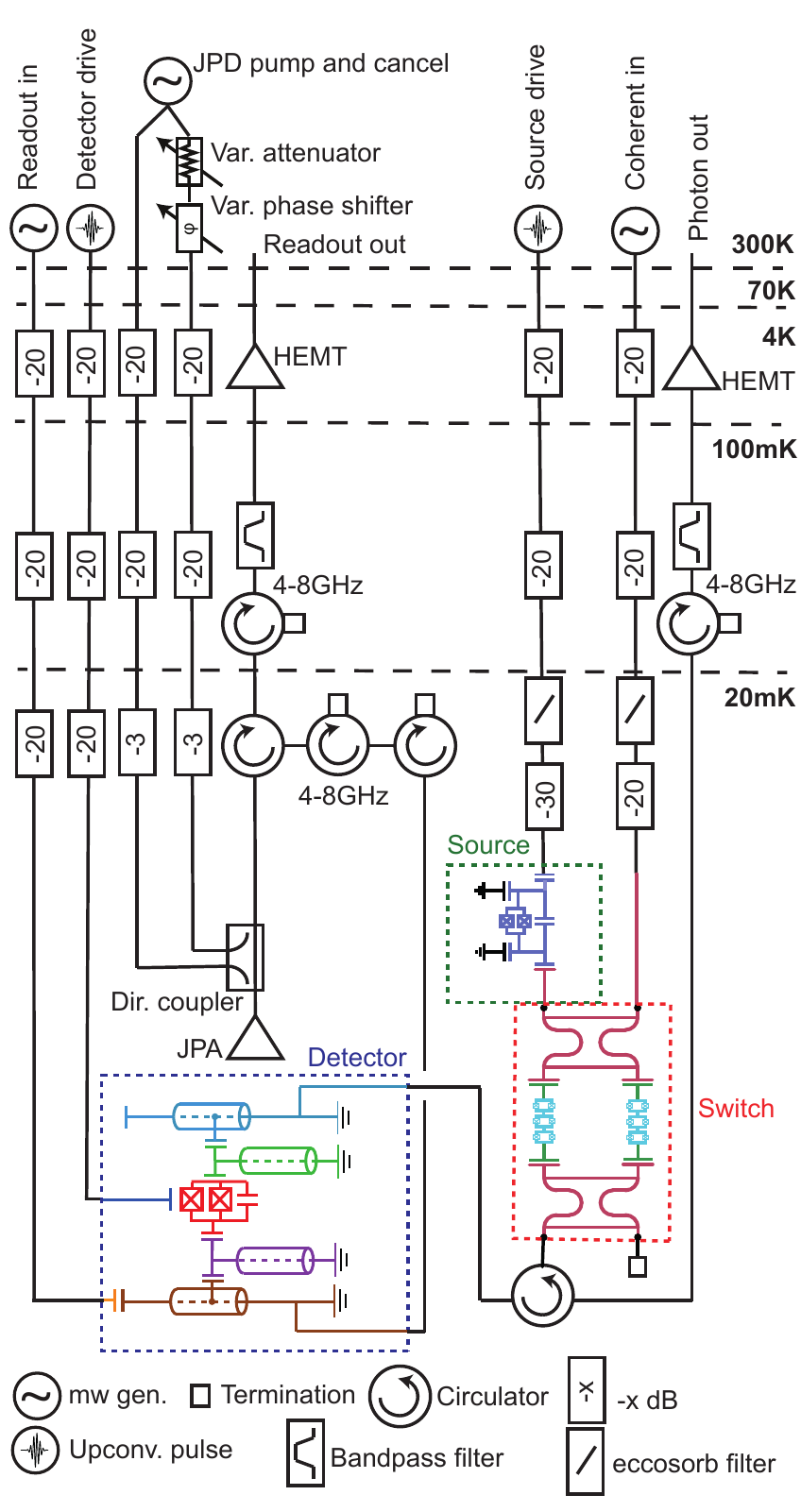}
\caption{\label{fig:setup} Scheme of the experimental set-up. All microwave lines are inner/outer DC-blocked at the fridge flange. Source and switch are physically on the same sample. DC cabling for applying external magnetic field with two coils on the switch sample holder and one coil at the detector are not shown.}
\end{figure}

The sample, shown in Fig.~\ref{fig:samples}, is fabricated on a 4.3~mm x 7~mm Sapphire chip cut along c-plane. All elements except for the qubit are fabricated from a 150~nm-thick sputtered niobium film using photolithography and reactive ion etching. The transmon's islands and Josephson junctions are fabricated in a second step using electron-beam lithography and shadow-evaporation of aluminum in an electron-beam evaporator. Both the photon detection device and the single photon source embedded in an on-chip switch~\cite{Pechal2016suppl} are mounted at the base temperature stage (20~mK) of a dilution refrigerator, as shown in the wiring diagram in Fig.~\ref{fig:setup}.

\subsection{Calibration of photon loss}
We distinguish internal detector inefficiencies from photon losses by measuring the attenuation constant between the single photon source and the detector. This measurement relies on operating both devices as calibrated power sources and comparing the relative power levels at room temperature.

\begin{figure}
\includegraphics{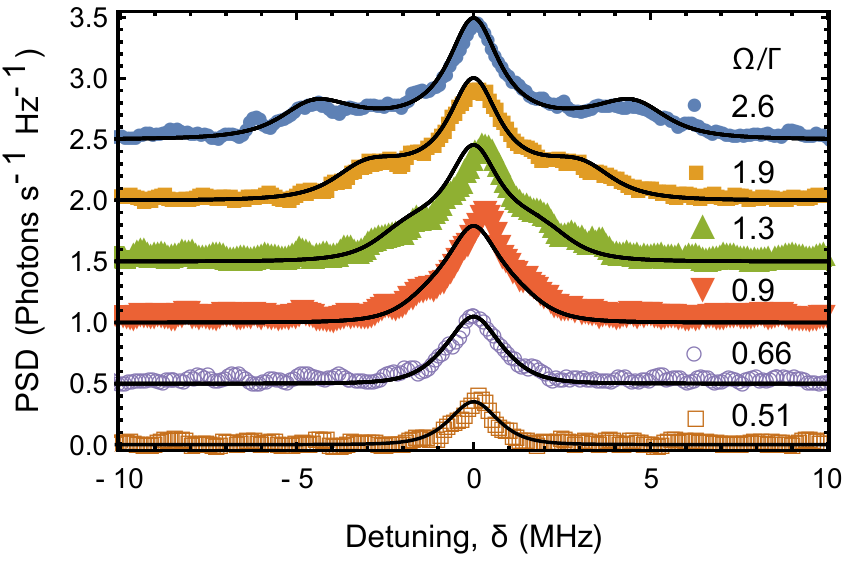}
\caption{\label{fig:Mollow} Measured power spectral density (PSD) of the inelastic scattering of a coherent tone resonant with the source-qubit (symbols) for various drive rates $\Omega$ in units of the linewidth $\Gamma$. Solid lines are fits to the data. Individual data sets are offset by 0.5~photons~s\textsuperscript{-1}Hz\textsuperscript{-1} for clarity.}
\end{figure}

First, we employ the photon-blockade effect of the photon source-qubit to realize a calibrated power source. We continuously drive the qubit at its transition frequency $\omega_\mathrm{ge}$ and measure the power spectral density of the inelastically scattered radiation emitted into the output port (Fig.~\ref{fig:Mollow}). In the limit of large drive rate $\Omega > \Gamma$ the measured spectrum features characteristic satellite peaks at detunings $\delta\approx\pm\Omega$ relative to the drive frequency~\cite{Lang2011suppl}. This nonlinear property of the spectrum allows one to calibrate the emitted power $P_\mathrm{s}=n_\mathrm{q}\Gamma\hbar\omega_{ge}$ from a global fit of the Mollow triplets. Here, $\Gamma/(2\pi)=1.77$~MHz is the source-qubit linewidth, and $n_\mathrm{q}$ the steady state average excited state population of the qubit. We note that in the limit of large drive rate $\Omega \gg \Gamma$, the qubit is driven into a mixed state with $n_\mathrm{q}\approx1/2$. Based on this fit we obtain
the ratio $G_\mathrm{s}$ of the power detected at room temperature and the power $P_\mathrm{s}$ emitted from the source. $G_\mathrm{s}=(1-L)G_d$  is composed of the photon loss $L$ from the source to the detector and the effective gain $G_d$ from the output of the photon detector to the room temperature electronics.

\begin{figure}
\includegraphics{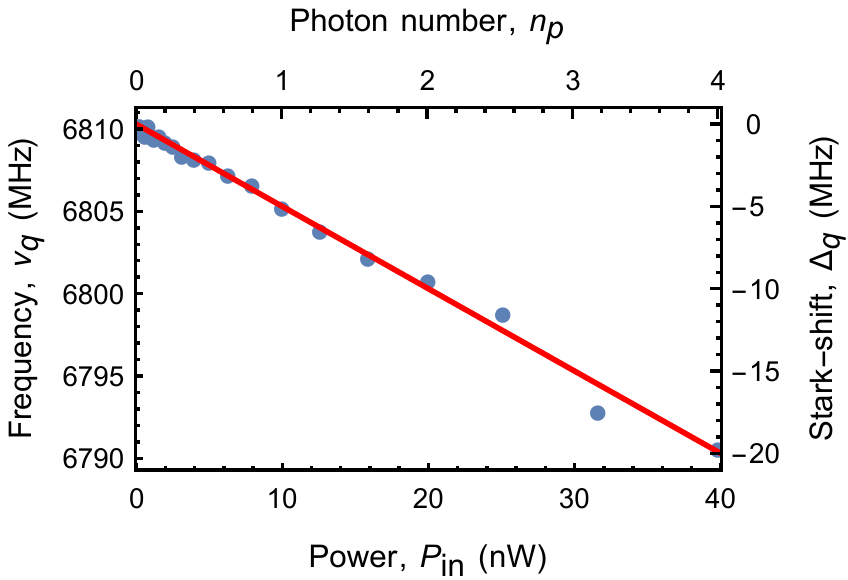}
\caption{\label{fig:ACStark} Stark-shifted frequency $\nu_\mathrm{q}$ of the detection-qubit (blue dots) as a function of the input power $P_\mathrm{in}$ at the generator (bottom axis), as well as the inferred photon number $n_\mathrm{p}$ in the cavity (top axis). The red curve is a linear fit to the data.}
\end{figure}

To calibrate the gain $G_d$, we operate the photon detector as a calibrated power source. For this purpose we tune the detection-qubit to its sweet spot, detuned from the detector resonator by $\Delta/(2\pi)=(\omega_\mathrm{cav}-\omega_\mathrm{ge})/(2\pi)=-676$~MHz, and populate the detector resonator using a coherent tone applied through the second port of the switch. We measure the power-dependent qubit frequency $\omega_\mathrm{q}$, which decreases linearly with applied power $P_\mathrm{in}$ (Fig.~\ref{fig:ACStark}). This is due to the AC Stark shift $\Delta_\mathrm{q}=\omega_\mathrm{q}-\omega_\mathrm{q}^0=2\chi n_p$~\cite{Schuster2005suppl}. Independently, we infer the dispersive shift $\chi/(2\pi)=\alpha g^2/(\Delta (\Delta-\alpha))/(2\pi)=-2.4$~MHz from spectroscopic measurements of the qubit anharmonicity $\alpha/(2\pi)=-340$~MHz, and the resonant qubit-cavity coupling $g_\mathrm{0}/(2\pi)=40$~MHz. This yields a calibration for the number of photons $n_\mathrm{p}$ in the detector resonator.
Knowing the effective linewidth $\kappa$  of the cavity we extract the expected photon power $P_\mathrm{d}=\kappa n_\mathrm{p} \hbar \omega_\mathrm{cav}$ at the output of the detector. A comparison with the power measured at room temperature yields the effective gain $G_d$ of the amplification chain. The loss between the source-qubit and the detector resonator is thus estimated as $L_\mathrm{s-d}=1-G_\mathrm{s}/G_\mathrm{d}=0.25$.

We attribute this loss to the following main contributions. First, the circulator placed between the two chips has an insertion loss specified by the manufacturer to be 8\%. Second, the finite directivity of the single-pole, double-throw switch~\cite{Pechal2016suppl} contributes to effective losses by routing about 5\% of the power to its second output, terminated by a 50~$\Omega$ load at base temperature. Third, the SMP connectors and female-female bullets used to couple the radiation from a printed circuit board (PCB) to a microwave cable also contribute to a manufacturer-specified insertion loss of 5\%. Finally, the attenuation in 50~cm of CC85Cu cables connecting the two samples via the circulator amounts to approximately 2\% loss~\cite{Kurpiers2017suppl}. The total identified sources of loss add up to approximately 20\%. The difference relative to the loss of 25\%, extracted from the power detection measurements, could be due to factors not accounted for such as impedance mismatches or internal losses of components along the detection path, in particular at wirebonds between samples and PCBs.

\subsection{Single-shot readout and detection}
\begin{figure}
\includegraphics{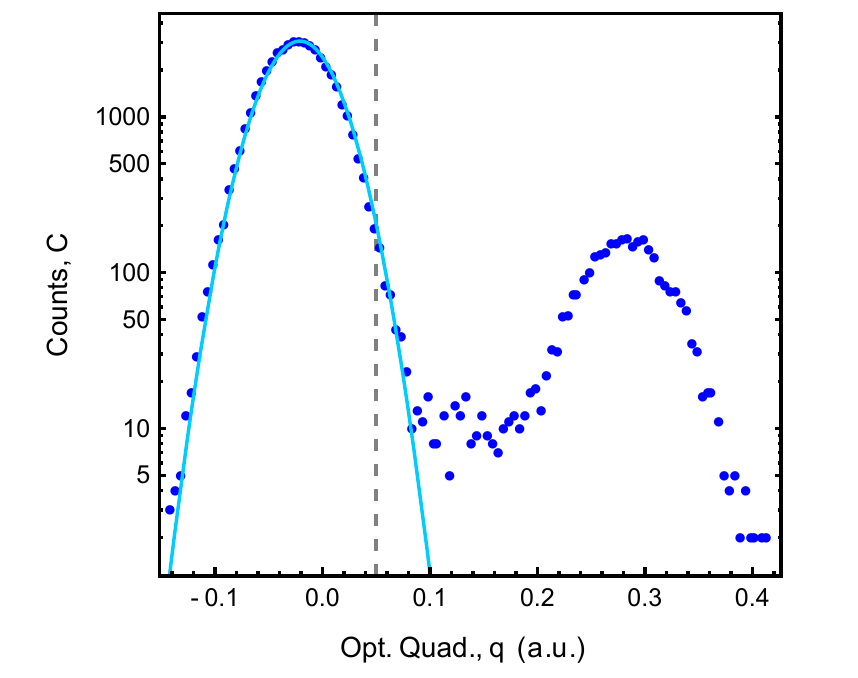}
\caption{\label{fig:preselect} Histograms of the integrated quadrature amplitude $q$ of the initial pre-selection measurement performed to reject initially excited qubit states. The solid line is a fit to a Gaussian. The dashed gray line indicates the chosen threshold. The 6\% of measurement traces above the threshold have been discarded.}
\end{figure}

Each experimental sequence starts with a measurement pulse, used to reject approximately 6\% of all measured traces (Fig.~\ref{fig:preselect}), in which the qubit was initially found in the excited state. Such instances are due to residual excitations from the previous run and  thermal excitations. To realize this pre-selection measurement we perform single-shot readout with the methods described in Ref.~\cite{Walter2017suppl}. After integrating the signal we obtain an integrated quadrature amplitude $q$ for each realization which we compare to a conservatively chosen threshold value to herald the qubit ground state.
 
\begin{figure}
\includegraphics{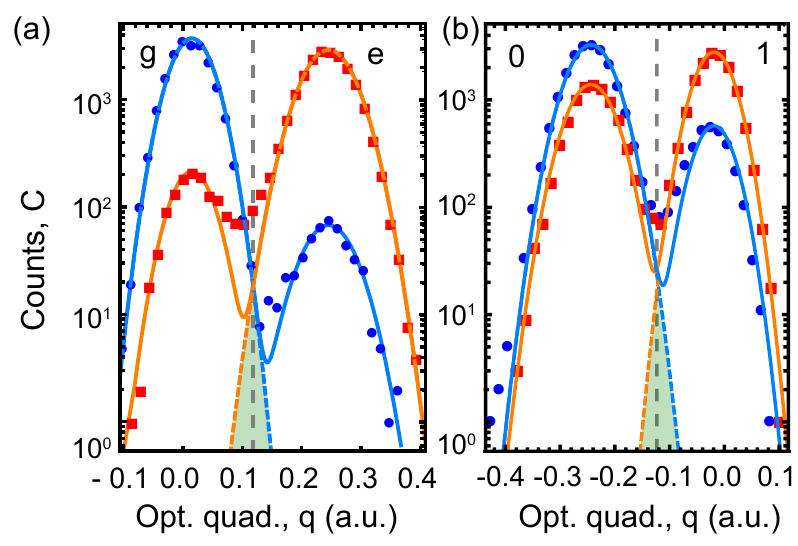}
\caption{\label{fig:shots} Single shots measurements. (a) Histogram of integrated quadrature amplitude $q$ for preparing the state $|g\rangle$ (blue) or $|e\rangle$ (red), as well as (b) for preparing the photon state $|0\rangle$ (blue) or $|1\rangle$ (red). In each panel the solid lines are fits to a double-Gaussian model whose individual components are indicated by dashed lines. The green areas depict the overlap error. The dashed gray line indicates the qubit-state threshold.}
\end{figure}

To characterize the qubit readout fidelity, we prepare 12,500 repetitions of each of the detection-qubit ground $|g\rangle$ and first excited states $|e\rangle$. We perform readout in 100~ns with a gated measurement pulse, obtaining the histograms shown in Fig.~\ref{fig:shots}a. A readout fidelity of $(91.5\pm0.3)\%$ is extracted. The errors are composed of $P(g|e)=(6.3\pm0.2)\%$ and $P(e|g)=(2.2\pm0.1)\%$. The overlap error is below 0.2\%.

To obtain the single-shot photon detection fidelity, we carry out the same readout procedure after 12,500 realizations of each emitting a single photon ($|\gamma\rangle=|1\rangle$) or not ($|\gamma\rangle=|0\rangle$), in both cases performing the detection protocol. The histograms are reported in Fig.~\ref{fig:shots}b. This corresponds to the single-shot single photon detection fidelity of $F=(49.6\pm0.5)\%$ reported in the main text. The infidelity is due to the finite detection efficiency, $P(g|1)=(37.0\pm0.4)\%$, dominated by losses between the source and the detector, and the dark count probability, $P(e|0)=(13.4\pm0.2)\%$, dominated by the detection-qubit decoherence during the length of detection window $T_\mathrm{w}$.

\begingroup
\def\refname{References}
\def\bibname{References}

\endgroup

\end{document}